\documentstyle[aps,prl,twocolumn]{revtex}
\input epsf

\def \be{\begin{equation}}
\def \ee{\end{equation}}

\def \twist{{\cal T}w}
\def \writhe{{\cal W}r}
\def \link {{\cal L}k}

\begin{document}
\title{Conformations of Circular DNA}
\author{Boris Fain and Joseph Rudnick}
\address{Department of Physics,
University of California, Los Angeles CA 90095, USA}
\maketitle
\begin{abstract}
We examine the conformations of a model for short closed 
DNA. The molecule is represented as a cylindrically 
symmetric elastic string with a constraint corresponding to a specification of the 
linking number. 
We obtain analytic expressions for configuration, elastic energy, 
twist and linking number for a family of solutions spanning from
a circle to a 'figure eight'. We suggest ways to use our construction
to make other configurations and models relevant to studies of
DNA structure. We estimate the effect of thermodynamic fluctuations.
\end{abstract}

\section{introduction}
The elastic model of DNA has been the subject of intense
research in the past 30 years. The approaches taken include
Lagrangian mechanics\cite{benham,lebret,tsu,tanak,manning},
 b-splines and (numerical) molecular
dynamics \cite{olsen}, and statistical mechanics\cite{siggia}.
 Until now no one has
been able to find the general equilibrium solutions for a
bent twisted segment of the molecule. As a particular case the
conformations of closed circular DNA have remained elusive.

In this paper we present a method for obtaining the shape,
twist, elastic energy and linking number of an isotropic
elastic segment subject to constraints. We will use this formalism
to describe unknoted circular DNA. We will also outline
how our formalism can be exploited to produce other shapes
of interest to DNA researchers.

\section{Elastic Model and Expressions of Interest}

In a previous work we outlined some developments in 
the elastic model of DNA\cite{fain}.
The molecule is represented as a slender cylindrical elastic
rod. At each point $s$ we describe the rod
by relating the local coordinate frame ${\cal L}$ to the frame ${\cal L}_0$
rigidly embedded in the curve in its relaxed configuration.
The relationship between the stressed and
unstressed local frames is specified by Euler angles $\theta(s),
\phi(s), \psi(s)$ needed to rotate ${\cal L}_0$ into $\cal L$.
Let us summarize some of the results needed for this paper.
Let the elastic constants of bending and torsional stiffness be
denoted,
respectively, by  $A$ and $C$, and let $L$ be the lengh of the
rod. The
elastic energy is given by
\be
E_{el}=\int_{0}^{L}ds\;\left(
\frac{A}{2}\left(\dot{\phi}^2\sin^2\theta+\dot{\theta}^2\right) +
\frac{C}{2}\left(\dot{\phi}\cos\theta+\dot{\psi}\right)^2\right)
\label{c_energy_eqn}
\ee

The twist is
\be
{\cal T}w=\frac{1}{2\pi}\int_{0}^{L}ds\;
\left(\dot{\phi}\cos\theta+\dot{\psi}\right) \label{tw_eqn}
\ee
A major relevant result from our previous work is that 
$\link$, the linking number, can be written as
\be
{\cal L}k\;=\;-1\;+\;\frac{1}{2\pi}\int_{0}^{L}\left(\dot{\phi}
+\dot{\psi}\right)\;ds \;\; \label{lk_eqn}
\ee
Where we have used Fuller's theorem\cite{fuller} to obtain the writhe,
and White's theorem\cite{white} for the total link. We shall use these
local quantities to determine the behaviour of the family of 
closed solutions that extremize the
elastic energy.

\section{Candidates: The Elastic Extrema}
\subsection{A Family of Curves}
We are searching for a family of solutions which are extrema
of elastic energy of a closed molecule. This family of 'writhing' solutions 
will be
indexed by the writhe, $\writhe$, of a member - ranging from $\writhe=0$ for 
a simple circle to $\writhe=1$ for a 'figure eight'. 
We know that for 
$\link \leq \sqrt{3}A/C$ the molecule remains stable as a circle
\cite{benham,boris}. 
 Therefore one bounding member of the family 
must be a simple circle, say, the $XY$ plane with a
 $\twist=\link=\sqrt{}3A/C$. 
We shall see that on the other end the bounding member, a 'figure eight', 
must have $\twist=0$ in order to be an extremum of elastic energy. 
However, the reader must remember that the actual constraint imposed 
on the closed curve is that of a fixed $\link$. We shall therefore examine 
the conditions under which the family of writhing curves can satisfy 
an imposition of a linking number constraint. 

\subsection{Euler-Lagrange Minimization}
We want to extremize elastic energy and keep the curve closed. 
We find that the functional to consider for
closed configurations of DNA is

\be
{\cal H} = \int_{0}^{L}dE_{el} - F{\bf t}{\bf{\hat z}}\:ds
 \label{c_hamil_eqn}
\ee

The first term of (\ref{c_hamil_eqn}) is clear - the elastic
energy must be extremized. The second term will enforce the
closure in the ${\bf \hat z}$ direction only because we can
always find a reference frame where the curve is closed in the
$XY$ plane.  Written explicitly as a function
of Euler angles (\ref{c_hamil_eqn}) becomes
\begin{eqnarray}
{\cal H} = \int_{0}^{L}\;ds \hspace{0.2in}
&&\frac{A}{2}\left(\dot{\phi}^2\sin^2\theta+\dot{\theta}^2\right)+ 
\nonumber \\
 +&&\frac{C}{2}\left(\dot{\phi}\cos\theta+\dot{\psi}\right)^2
 -F\cos\theta  \label{c_hamiltonian}
\end{eqnarray}
The extrema are found by applying Euler-Lagrange equations
to (\ref{c_hamiltonian}). Denoting the conserved quantities as
$p_{\phi} \equiv \frac{\partial{\cal H}}{\partial \dot \phi}$
and $p_{\psi} \equiv \frac{\partial{\cal H}}{\partial \dot \psi}$
we get
\begin{eqnarray}
\dot\phi&=&\frac{p_\phi-p_\psi\cos\theta}{A\sin\theta}
\label{c_phi_eqn}\\
\dot\psi&=&\frac{p_\psi}{C}-\dot\phi\cos\theta
\label{c_psi_eqn}
\end{eqnarray}
The equation for $\theta$ is a quadrature obtained by
integrating $\frac{\partial{\cal H}}{\partial \theta}
=\frac{d}{ds} \frac{\partial{\cal H}}{\partial \dot \theta}$
 with $E_0$ as the constant of integration.
Defining $u\equiv\cos\theta$ we see that the behavior of
solutions is governed by a cubic polynomial in $u$:
\begin{eqnarray}
{\dot u}^2&=&\frac{2\left(1-u^2\right)}{A}
\left(E_0-Fu\right)-
\frac{1}{A^2}\left(p_\phi^2+p_\psi^2-2 p_\phi p_\psi u \right)
\nonumber \\
&\equiv&\frac{2F}{A}\left(u-a\right)\left(u-b\right)\left(u-c\right)
\label{theta_eqn}\\
&&\mbox{where we order the roots }c\leq u(\equiv\cos\theta)\leq b\leq a
\nonumber
\end{eqnarray}
(\ref{theta_eqn}) requires $\cos\theta$ to oscillate between $c$ and $b$.
For a circular solution the tangent $\bf t$ will go back and forth twice.

\begin{figure}
\begin{center}
\leavevmode
\epsfxsize=3in \epsfbox{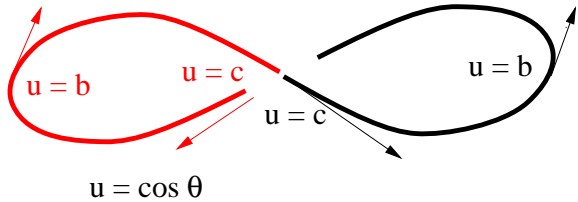}
\end{center}

\caption[]{\label{segment_fig}
The tangent oscillates $c\rightarrow b\rightarrow c\rightarrow b
\rightarrow c$.
 The curve is composed
of two symmetric parts.}
\end{figure}

All the relevant quantities, including the shape of the curve,
can be obtained using (\ref{theta_eqn}). For example:
\begin{eqnarray}
\phi(s)&=&\int_{u(s=0)}^{u(s)}\frac{d\phi}{ds}\frac{ds}{du}du
\;\;\;\mbox{ with } \nonumber\\
\frac{ds}{du}&=&\sqrt{\frac{A}{2F}}
\frac{1}{\sqrt{\left(u-a\right)\left(u-b\right)\left(u-c\right)}}
\label{c_quadrature_eqn}
\end{eqnarray}

\section{Constraints and Parameters}
\subsection{Constraints}
To complete the analysis we must determine the parameters generated 
by the minimization procedure: the invariants $p_\phi$ and 
$p_\psi$, the constant $E_0$, and the Lagrange multiplier $F$. 
To determine the parameters we impose constraints on the curve.
Figure \ref{segment_fig} is a good visual guide to the geometric
meaning of the constraints that follow:
\begin{eqnarray}
&\int_{0}^{\frac{L}{2}}ds=L/2 \;\;\;\;
&\mbox{(fixing the length at $L$)} \label{l_constraint}\\
&\int_{0}^{\frac{L}{2}}u(s)ds=0 \;\;\;\;
&\mbox{(enforcing closure in $z$)}\label{z_constraint}\\
&\int_{0}^{\frac{L}{2}}\phi(s)ds=\pi\;\;\;\;
&\mbox{($\phi$ must go all the way around)}\label{phi_constraint}
\end{eqnarray}

\subsection{Reparametrization}
At this point we reparametrize the problem by $F,a,b,c$ instead of
$F, E_0, p_\phi, p_\psi$. Parametrizing the problem by the roots of 
the polynomial is extremely advantageous: 
it makes the analytic manipulations more transparent; it
also streamlines the computational tasks. The two sets of
parameters are related in the following manner:
\begin{eqnarray}
E_0&=&F\left(a+b+c\right))\\
p_\phi&=&\sqrt{\frac{AF}{2}}\left( p_1 \pm p_2\right);\;\;\;
p_\psi=\sqrt{\frac{AF}{2}}\left( p_1 \mp p_2\right)\\
&\mbox{with}&\;p_{1\choose 2}\equiv\left[(c \pm 1)(b \pm 1)(a \pm 1)\right]^{1/2}
\nonumber
\end{eqnarray}
The choice of branch ($\pm$) is imposed by the family of configurations sought.
For circular DNA without intrinsic curvature, $p_\phi$ takes a $-$,
and $p_\psi$ correspondingly a $+$.\footnote{the choice of a particular branch is
a non-trivial procedure} Let us make some definitions that make notation more 
transparent: 
\begin{eqnarray} 
&&\Delta b\equiv \left(b-c\right)\;\; ; \;\; \Delta a\equiv \left(a-c\right) \label{delta_eqn}\\
&&q\equiv\sqrt{\frac{\Delta b}{\Delta a}} \label{q_eqn}
\end{eqnarray}
Employing (\ref{c_quadrature_eqn})
we rewrite the constraint equations,
(\ref{l_constraint}), (\ref{z_constraint}),
and (\ref{phi_constraint}), in terms of the new parameters:
\begin{eqnarray}
&&F=\frac{32A}{\Delta a L^2}K\left(q\right)\;\;\;\;
\label{c_f_constraint}\\
&&aK\left(q\right)=\Delta aE\left( q \right)
\label{c_a_constraint}\\
&&\pi/2\;\;=\;\;\frac{1}{\Delta a}\left[
\left|\frac{(a+1)(b+1)}{(c+1)}\right|^{1/2}
\Pi\left(\frac{\Delta b}{-1-c},q \right)\right.- \nonumber \\
&&\hspace{2.15cm}-\left.\left|\frac{(a-1)(b-1)}{(c-1)}\right|^{1/2}
\Pi\left(\frac{\Delta b}{1-c},q \right)\right]
\label{c_phi_constraint}
\end{eqnarray}
Where $K,E\mbox{ and }\Pi$ are complete elliptic integrals of the first,
second and third kind, respectively.

\subsection{Solving for $a$, $b$, $c$ and $F$}

We find that the optimal procedure for calculation of solutions' parameters 
is to use (\ref{c_f_constraint}) to eliminate $F$, then
(\ref{c_a_constraint})
and (\ref{c_phi_constraint}) eliminate $\Delta a$ and $c$. Because 
the $\writhe$ is a monotonically increasing function of 
$\Delta b$, we leave it to classify the family of
curves. Once the constants $F,a,b,c$ are determined, the desired
solutions and all the relevant quantities are computed via
elliptic integrals. For example, the explicit expression for
$\theta(s)$ in the first quarter of oscillation is (we have inverted (\ref{c_quadrature_eqn})):
\be
\cos\theta=\Delta b \mbox{ sn}^2\left(\sqrt{\frac{F\Delta c}{2A}}s,
\sqrt{\frac{\Delta b}{\Delta a}}\right)-c \label{c_explicit_theta_eqn}
\ee
$\phi \mbox{ and }\psi$ are obtained similarly from
(\ref{c_phi_eqn}) and (\ref{c_psi_eqn}).

\noindent The following figure shows the family of curves
computed in the
manner discussed above. Since the constraint equations involve
elliptic integrals, finding a solution on a computer is virtually
instanteneous.\footnote{analytically and computationally elliptic
integrals are equivalent to, say, $\arcsin$}

\begin{figure}
\begin{center}
\leavevmode
\epsfysize=2in \epsfbox{c1.cps}
\epsfysize=2in \epsfbox{c2.cps}
\epsfysize=2in \epsfbox{c3.cps}
\end{center}
\end{figure}

\begin{figure}
\begin{center}
\leavevmode
\epsfxsize=3in \epsfbox{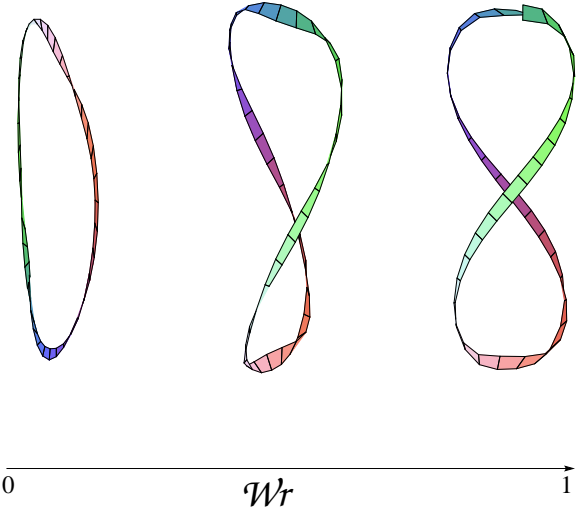}
\caption[]{\label{circle_family_fig}
The family of curves ranges from the circle in the $XY$
plane to the 'figure eight' in the $YZ$ plane. The perspective
is slightly assymetric to aid visualisation.}
\end{center}
\end{figure}

\subsection{Bounding Members: Circle and 'Figure Eight'}
Let us check whether the initial member of our family, a 
circle with $\writhe=0$ joins smoothly with the previously 
known stable family of twisted circles\cite{benham_stab,boris}.
The circle corresponds to $\writhe=0$.
A circle in the $XY$ plane the curve must have
$b_0=c_0=0$. (\ref{c_phi_constraint}) now states that 

\begin{eqnarray}
&&\frac{1}{\sqrt a_0}\left(\sqrt{a_0+1}-\sqrt{a_0-1}\right)=1\\
&&\mbox{which in turn gives}\nonumber\\
&&{p_\psi}_0=\frac{2\pi A}{L}\frac{1}{\sqrt a_0}
\left(\sqrt{a_0+1}+\sqrt{a_0-1}\right)=\frac{2 \pi A}{L}\sqrt{3}
\label{p_circle_equation}
\end{eqnarray}
Combining (\ref{tw_eqn}) and (\ref{c_psi_eqn}) to obtain the $\twist$ of the circle 
(\ref{p_circle_equation}) gives
\begin{equation}
\twist_0=\frac{L}{2\pi}{p_\psi}_0=\sqrt{3}\frac{A}{C}
\end{equation}

It is also of interest to compute the $\twist$ of the 'figure eight' 
which, like the circle, can be done virtually by inspection. The figure 
lies in the $YZ$ plane, which forces $\phi$ to behave as follows: (refer 
to Figure \ref{segment_fig} for visualization)
\begin{equation}
{\dot{\phi}}_8=\pi\left(\delta\left(0\right)+\delta\left(\frac{L}{2}\right)\right)
\label{c_phi_eight_eqn}
\end{equation}
Combining (\ref{c_phi_eight_eqn}) and (\ref{c_phi_eqn}) 
forces ${p_\phi}_8={p_\psi}_8=0$ which immediately sets 
$\twist_8=0$. The value of $\Delta b_8$ is easily determined 
from the fact that $c_8=-1$ (this can be seen from the curve itself) 
necessitating $a=1$. Then (\ref{c_phi_constraint}) gives 
$\Delta b_8=1.6522\ldots$.

\section{Linking Number and the Plectonemic Transition}
We have found a family of writhing solutions that are the extrema 
of elastic energy. The writhe of the curves covers the range $0\leq\writhe\leq1$.
\footnote{That the 'figure eight' has $\writhe=1$ just before crossing 
can be seen just from the shape of the curve.}
However, the physical constraint imposed on the molecule is 
$\link$, the linking number. The explicit expressions for $\twist$, $\link$ 
and $\writhe$ are easily obtained: 
\begin{eqnarray}
\twist&=&\frac{2\pi}{\sqrt{\Delta a}}K\left(q\right)\left(p_1+p_2\right)\frac{A}{C}\\
\writhe&=&\frac{2\pi}{\sqrt{\Delta a}}\left[-K\left(q\right)\left(p_1+p_2\right)\right.\nonumber\\
&+&\left.\frac{p_1}{1+c}\Pi\left(\frac{\Delta b}{-1-c},q \right)+
\frac{p_2}{1-c}\Pi\left(\frac{\Delta b}{1-c},q \right) \right]\\
&&\mbox{and, using White's theorem\cite{white}}\nonumber \\
\link&=&\twist+\writhe
\end{eqnarray}
Plotting $\link$ vs. $\writhe$ for our family we can discern three 
distinct types of behavior depending on the value of the ratio $A/C$. 

\begin{figure}
\begin{center}
\leavevmode
\epsfxsize=3in \epsfbox{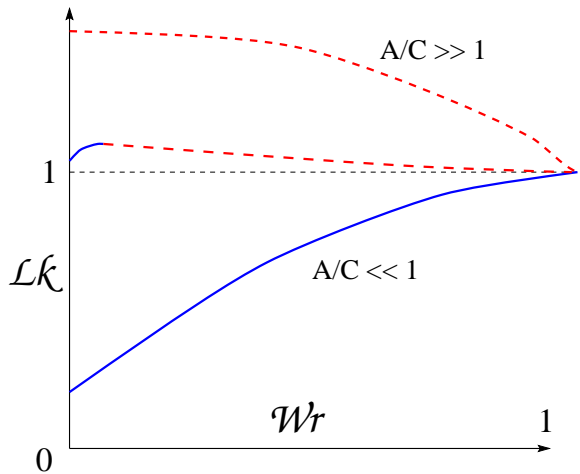}
\end{center}
\caption[]{\label{link_family_fig}
The three types of behaviour of $\link$ for our writhing 
family of curves. 
The solid line indicates conformations realized by the molecule, 
whereas the dashed regions are forbidden by the imposition of $\link$. 
The ratio $A/C$ controlls the plectonemic transition.}
\end{figure}

We can now describe what happens to the molecule as the linking number 
is increased. Initially the molecule remains a circle until 
$\Delta \link=\sqrt{3}A/C$. After that there are three possibilities. 
If $A/C\ll 1$ then the writhing family can support a steady increase 
to $\link=1$ of the 'figure eight' and the molecule folds continuosly 
until self-crossing. (Futher increase of $\link$ presumably results 
in a plectonemic configuration). If $A/C\gg1$ then none of the members 
of the writhing family can support the neccesary linking number, and as 
soon as $\Delta \link$ exceeds $\sqrt{3}A/C$ the twisted circle snaps 
into a plectoneme. An interesting case is the intermediate one, where, 
at first, the molecule starts writhing, but at some point, unable to support 
further increase in $\Delta \link$ it snaps into a plectoneme. Another 
interesting fact is that this behavior is {\em independent} of the 
length of the molecule. 

\section{Possible Generalizations}
\subsection{Intrinsic Curvature}
The method of construction of the circular DNA can be
easily generalized to other configurations of interest. Considerable
activity has been generated around the studies of DNA with
intrinsic curvature. We feel such generalizations are important 
because various chemical effects - hydrophobic, etc. - 
can be modeled by intrinsic curvature\cite{boles,bauer,tobias}. 
Inclusion of intrinsic curvature into
the present framework involves a simple modification
of the LHS of equation (\ref{c_phi_constraint})\cite{boris}. 
If the inital curvature is such that
the DNA is fully relaxed in the circular state, the characteristic
polynomial (\ref{theta_eqn}) degenerates into a quadratic and
all the quantities are expressible in terms of simple trigonometric
functions. \footnote{In the current formalism
(\ref{c_phi_constraint})will force $a\rightarrow\inf$; but it is
easier to omit the term $F\;\cos\theta$ in the initial functional
and derive the equations anew.} One complication does arise in
cases of initial curvature: the solutions have to be minimized
wrt to the initial value of $\psi$. An intuitive example of this
caveat is the fact that the energy of intrinsically curved loops
is changed by 'rolling', unlike that of the configurations which
are straight in their relaxed state.
\subsection{Trefoils and Other Torus Knots}
Another interesting generalization of our results is obtained by 
altering the closure condition\cite{boris,zohar}. For example, 
a curious configuration is attained by requiring the curve
to close after $\phi$  has turned $4\pi \mbox{ rather than the }
2\pi$ required for the closure of a circular conformation.
The resulting curve is a trefoil knot. These curious shapes are 
probably the analytic analogs of similar configurations observed 
by Shlick and Olsen\cite{olsen}.
{\em cite myself unpublished}
\section{Entropic Corrections}

Although our model is applicable to the cases when thermodanamic
fluctuations are not important (i.e. short segments) it is worth
while to estimate the contribution of thermal fluctuations to the
free energy. Since the aim of this paper is to derive the elastic
shapes of DNA, we will concentrate on thermal effects that
influence curvature. This can be done in analogy to Marko and Siggia
\cite{siggia_science} who, in turn, used the scaling
ideas originally developed for the study of fluctuating membranes\cite{safran}.
Their (slightly generalized) dimensional analysis states that the 
entropic contribution to the free energy can be written as 
\begin{eqnarray}
{\Delta f}& =&k_BT\:\left(\frac{\kappa^2}{A/k_BT}\right)^{1/3}
\label{c_entropy_eqn}\\
&&\mbox{where the curvature $\kappa$ is} \nonumber \\
\kappa^2&=&\dot\theta^2+\dot\phi^2\sin\theta^2
\end{eqnarray}
We conceed that equation (\ref{c_entropy_eqn}) gives only the
form of the thermal correction, sans the multiplicative factor
of order unity. (\ref{c_entropy_eqn}) reflects the fact that increasing
the curvature reduces the correlation length, thus suppressing
fluctuation and incurring a corresponding loss of entropy.

\section{Conclusion}
We have presented a formalism for obtaining the elastic mimima of circular 
DNA subject to a constraint in the linking number. Aside from the simplicity 
of the statement of the problem, and the beauty (in the eyes of the creators) 
of the solution, our formalism can significantly improve the investigations 
of linear molecules with elastic models. This method can be 
easily generalized to create a truly 'finite' finite-element-analysis in which 
elastic models can be traced out by finite segments. The conformations 
of these segments can be specified in close form which would decrease 
computation time significantly.

\end{document}